\documentclass[showpacs,superscriptaddress,floatfix,amsmath,amssymb,twocolumn,pre,longbibliography]{revtex4-1}
\usepackage{graphics} 
\usepackage{graphicx}
\usepackage{epsfig}
\usepackage{amsmath}
\usepackage{amsfonts}
\usepackage{bm}
\usepackage{color}
\def\vx{\bm x}
\def\vy{\bm y}

\newcommand{\xib}{{\bm \xi}}

\def\bx{\mathbf{x}}
\def\by{\mathbf{y}}
\def\rd{{\rm d}}
\def\H{\mathcal{H}}

\date{}

\begin{document}
\title{{Asymmetric noise-induced large fluctuations  in coupled systems}}

\author{Ira B.~Schwartz$^1$ and Klimka Szwaykowska}

\affiliation{U.S. Naval Research Laboratory\\ Code 6792, Plasma Physics Division,
  Nonlinear Systems Dynamics Section\\ Washington, D.C., 20375,
  USA\\email:Ira.Schwartz@nrl.navy.mil\\tel:202-404-8359 fax:202-767-0631}
\author{Thomas W. Carr}
\affiliation{Department of Mathematics, Southern Methodist University, Dallas, Texas 75275, USA}


\date{\today}
\begin{abstract}

Networks of interacting, communicating subsystems are common in many
fields, from ecology, biology, epidemiology to engineering and  robotics. In
the presence of noise and uncertainty, interactions between the individual
components can lead to unexpected complex system-wide behaviors. In this
paper, we consider a generic model of two weakly coupled dynamical systems,
and show how noise in one part of the system is transmitted through the
coupling interface. Working synergistically with the coupling, the noise on
one system drives a large fluctuation in the other, even when  there is no
noise in the second system. Moreover, the large fluctuation happens while the
first system exhibits only small random oscillations.  Uncertainty effects are
quantified by showing how characteristic time scales of noise induced switching scale as a function of  the coupling between the two coupled parts of the experiment. In addition, our results show that the probability of switching in the noise-free system scales inversely as the square of reduced noise intensity amplitude, rendering the virtual probability of switching to be an {\it extremely rare event.} Our results showing the interplay between transmitted noise and coupling are also  confirmed through simulations, which agree quite well with analytic theory.

\end{abstract}
\maketitle




\section{Introduction}

Understanding the interaction between noise and system dynamics is key to
  understanding unexpected system behaviors~\cite{gar03,vanKampen_book}, and hence to
 robust and efficient operation of autonomous systems deployed in noisy, uncertain
  environments. It is often assumed that dynamics with small noise input can
  be modeled as small perturbations of the deterministic system dynamics;
  however, there are many known cases where small noise inputs can drive
  large-scale transitions in system behavior. Examples include noise-induced
  switching between attractors in continuous systems
  \cite{FW84,Doering1992,Castellano1996,Lapidus1999,Kim2005,Siddiqi2005,Chan2007,DS2012,Dhuys2014a,Emen2016,Huang2016},
  and noise-induced switching and extinction in finite-size systems
  \cite{Allen2005,Barzel2008a,Kamenev2008b,Meerson2008,Dykman,Wells2015,Chen2016a,Hindes2016}.
  
In both  switching and extinction, a significant change in the state of the
  system occurs as the result of a noise-induced large fluctuation. For
  systems with small noise, such a large fluctuation is a rare event,
  and occurs on average when the noise signal lies along a
  so-called ``optimal path''~\cite{schwartz2011converging}. For systems
  operating in most common environments, noise is assumed to be homogeneous,
  and it is relatively straight-forward to compute the optimal paths which
  lead to large fluctuations~\cite{Dykman1990}.

 In contrast to homogeneous noise, finite systems, whether
     continuous or discrete, are often subject to asymmetric noise~\cite{ISI:000263389500058,ISI:000082402400005}. One excellent
   example of multiple independent noise sources occurs in coupled finite communicating systems
     operating in noisy environments~\cite{Lindley2013a}, where the effects of noise on the collective motions
   of swarms of self-propelled autonomous agents results in drastic pattern
   changes. Such systems are of tremendous practical importance; coordinated
   groups of agents have been deployed for a wide range of applications,
   including exploration and mapping of unknown environments
   \cite{Earon2001,Cheng2011,Bhattacharya2012,Lynch2008,Wu2011}, search and
   rescue \cite{Takano2014,AlTair2015}, and construction
   \cite{Augugliaro2014}. Extensions to the basic swarming dynamics by using
   teams of heterogeneous agents capable of cooperatively executing more
   complex tasks are presented in \cite{Ramp2009,Dorigo2013}. In addition,
   network structure and uncertainties in delay communication have been shown
   to give rise to dynamic patterns in collective swarm motion
   \cite{szwaykowska2016collective,hindes2016hybrid}.

   Usually, sophisticated
   models are used to predict  behavioral patterns for large groups of
   interacting individual agents
   \cite{b-jccvg97,Calovi2014,Carlen2013,Hsieh2008a}. However, testing these
   behaviors in real-world environments often presents significant logistical
   challenges. In many cases, it is more practical to rely
   on mixed-reality experiments (similar to ideas
   in~\cite{Gintautas2007}), where real agents are deployed
   alongside simulated ones, in order to better understand how real-world
   noise affects the collective dynamics, as well as validate the theory
   against a critical number of agents~\cite{Szwaykowska2016}. This creates a situation where we have two coupled systems with asymmetric noise characteristics: the set of real agents, operating in a high-noise real-world environment, and the simulated agents, operating in a (at least partly) idealized simulated world. Our current paper is inspired by this situation; we consider a generic pair of coupled dynamical systems, and study the effects of interaction on switching in the low-noise system.

As shown e.g. in~\cite{Kozlov2016}, even weak coupling between system
   dynamics can significantly affect the behavior of the coupled systems.  We
   show that even weak interaction between a low noise, or noise-free,
   simulated system and a noisy ``real'' system can cause catastrophic
   transitions between states. That is to say, even if only part of the system
   operates in noisy real-world conditions, we can observe large changes in
   the dynamics of the idealized, low-noise virtual part, since noise is
   transmitted from the real to virtual world via coupling. Since one of our
   main results shows how the probability exponent is enhanced by the ratio of
   two noise sources, we refer to the state transitions induced in the noise-free virtual system by coupling with the noisy real-world system as  extreme rare events.

 The rest of the paper is laid out as follows. In
 Section~\ref{sec:ProblemSetupInGeneral} we define the general asymmetric
 noise problem for coupled systems (which include MR systems). Gaussian noise
 is considered here, but the theory can be made more general to include
 non-Gaussian perturbations~\cite{Dykman_prefactor} and
 correlations~\cite{feyhib65,Dykman1990}. Noise induced large fluctuations are
 posed in a variational setting for the coupled problem. Linear response to
 the noise is derived in general. In
 Section~\ref{sec:AModelExampleOfMixedRealityNoiseInducedPerturbations}, we
 consider a model problem of coupled bi-stable attractors subjected to
 asymmetric noise. For the specific problem we compare our theory to Monte
 Carlo simulations, and derive scalings as a function of the heterogeneity of
 the noise. We derive a general scaling relation between the noise ratio and
 the coupling strength that governs the mean probability to switch. To
 quantify the extreme rare event of the low-noise system switching, we derive
 the exponent of the probability distribution, and show that this exponent varies as the inverse noise ratio squared. Comparisons between our general theory and numerical experiments of such large fluctuation events show excellent agreement. A discussion of the results and conclusions are given in Section~\ref{sec:DiscussionAndConclusions}.

\section{Problem Setup in General}
\label{sec:ProblemSetupInGeneral}

\begin{figure}[htb]
 \centering
 \vspace{5mm}
 \includegraphics[width=0.4\textwidth]{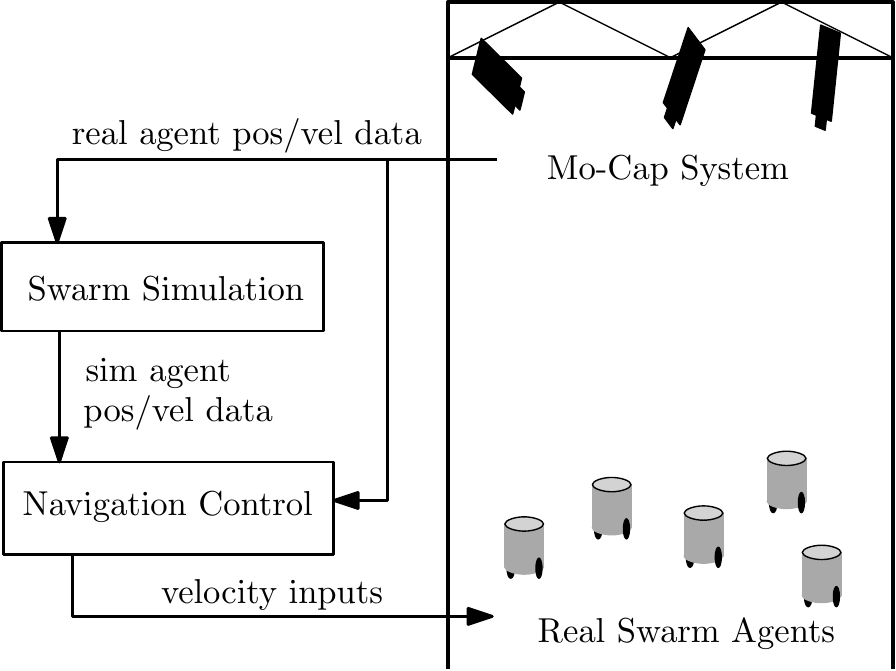}
 \caption{Experimental setup with virtual swarm of agents. The real robots operate in a lab testbed.  The positions of the real and simulated agents are passed to the
 virtual swarm simulator, which models the response of the virtual swarm agents to the current swarm configuration; and
 to the controller, which computes the real robot response and passes target
 velocities to the real swarm agents. The agents use internal proportional $-$
 integral $-$ derivative (PID) controllers to achieve the target velocities.}
 \label{fig:expsetup}
\end{figure}

The problem formulation described here is motivated by the mixed-reality system shown in Fig.~\ref{fig:expsetup}.  In this setup, the physical agents operate in an uncertain, noise-ridden environment, which imposes a larger noise source on all of the agents. 
In contrast,  the virtual agents are isolated from any real environmental perturbations, and experience only the noise modeled in the simulation.
We let the time dependent vectors ${\vx,\vy}$
denote the state-space configurations of agents operating in virtual and real environments, respectively.

We wish to examine the situation where there is a significant asymmetry in the noise characteristics of the two coupled systems; in particular, where the noise intensity in the low-noise (``virtual'') system goes to zero. 

\subsection{The stochastic equations of motion}
To analyze how noise impacts the dynamics from one environment to another, we consider a general coupled stochastic differential equation of the form
\begin{subequations} \label{eq:SDE1and2}
  \begin{align}
    \dot{\vx}(t) &= \boldsymbol{f}(\vx(t)) + {\bf h}_1(\vx(t),\vy(t),K)+
    \epsilon \boldsymbol{G}_1({\vx}(t))\boldsymbol{\xi_x}(t),
    \label{eq:SDE1} \\
    \dot{\vy}(t) &= \boldsymbol{f}(\vy(t)) +  {\bf h}_2(\vx(t),\vy(t),K) +
    \boldsymbol{G}_2(\vy(t))\boldsymbol{\xi_y}(t),
    \label{eq:SDE2}
  \end{align}
\end{subequations}
where $\boldsymbol{x}\in\mathbb{R}^{n_x}$ and $\boldsymbol{y} \in \mathbb{R}^{n_y}$ represent the state-space configurations of the low- and high-noise systems, respectively, and
  the matrices ${\boldsymbol G}_i$, $i=1,2$ \footnote{Throughout the paper, boldface lower-case letters will indicate vectors, while boldface upper-case letters will indicate matrices.}  are given by
$\boldsymbol{G}_i(\vx(t))=\text{diag}\{g_{i1}(\vx(t)),g_{i2}(\vx(t)),...g_{in_i}(\vx(t))\}$,
where the $g_{ij}$'s are general nonlinear functions. Coupling strength is
denoted by parameter $K$, and we choose ${\bf h}_1$, ${\bf h}_2$ so that ${\bf h}_1(\vx(t),\vy(t),0)={\bf h}_2(\vx(t),\vy(t),0) = 0$; 
i.e., the systems $\vx, \vy$ are uncoupled when $K=0$.

We assume that the noise inputs $\boldsymbol{\xi}_x\in\mathbb{R}^{n_x}$ and $\boldsymbol{\xi}_y\in\mathbb{R}^{n_y}$ are  independent Gaussian-distributed stochastic processes with independent components, and intensity $D$. They are both characterized by a probability density functional $\mathcal{P}_{\boldsymbol{\xi}}=e^{-\mathcal{R}_{\xi}/D}$, where $\mathcal{R}_{\xi}$ is defined as
\begin{equation}
\label{eq:Gauss_Functional}
{\cal R}_{\xib}[\xib(t)]= \frac{1}{4}\int dt\,dt'\,\xib(t)\xib(t').
\end{equation}

In order to capture the asymmetric noise levels between the two systems, we introduce a parameter, $\epsilon \ll  1$, that controls the noise
intensity of the state variable ${\vx}$.  The case $\epsilon = 0$ corresponds
to noise-free operation. However, even with $\epsilon = 0$, noise-induced transitions can still occur as a result of noise transference through the coupling with the high-noise system.

\subsection{Deterministic dynamics}

In the absence of any noise, Eqs.~\ref{eq:SDE1and2} are ordinary
differential equations, and we suppose that there exist steady states which
depend on the coupling strength, $K$. We therefore assume that there
exists an attracting equilibrium $({\vx}_a(K),{\vy}_a(K))$ and at least one
saddle equilibrium point, $({\vx}_s(K),{\vy}_s(K))$.

The stationary states satisfy.
\begin{subequations} \label{ssvectorfield}
  \begin{align}
    \boldsymbol{f}(\vx_a) + {\bf h}_1(\vx_a,\vy_a,K) &= \boldsymbol{f}(\vx_s) + {\bf h}_1(\vx_s,\vy_s,K)=0\\
    \boldsymbol{f}(\vy_a) +  {\bf h}_2(\vx_a,\vy_a,K) &= \boldsymbol{f}(\vy_s) +  {\bf h}_2(\vx_s,\vy_s,K)=0.
  \end{align}
\end{subequations}

The stability of the equilibrium states is given by the linear variational
equations of motion about that state:
\begin{equation}
\dot{\bf X}(t) = {\bf {\cal M}}({\bar \vx},{\bar \vy},K) {\bf X}(t),
\end{equation}
where ${\bar \vx},{\bar \vy}$ denote either $\vx_a,\vy_a$ or $\vx_s,\vy_s$, and
\begin{equation*}
{\bf {\cal M}}({\bar \vx},{\bar \vy},K) = 
\begin{bmatrix}
\frac{\partial {\bf f}({\bar \vx})}{\partial x} +  \frac{\partial {\bf h}_1({\bar \vx},{\bar \vy},K)}{\partial x} 
& \frac{\partial {\bf h}_1({\bar \vx},{\bar \vy},K)}{\partial y}\\
\frac{\partial {\bf h}_2({\bar \vx},{\bar \vy},K)}{\partial x}  &  \frac{\partial {\bf f}({\bar \vy})}{\partial y} +  
\frac{\partial {\bf h}_2({\bar \vx},{\bar \vy},K)}{\partial y} 
\end{bmatrix}.
\end{equation*}
The matrix ${\bf {\cal M}}({\bar \vx},{\bar \vy},K)$ evaluated at the  saddle
point is assumed to have only 
one positive real eigenvalue (associated with an unstable direction in the space
of dynamical variables), while the rest of the eigenvalue spectrum lies in the
left hand side of the complex plane. In particular, we assume that for all values of interest $K$, the saddle point
 lies on the basin boundary of the attractor. The generic switching
   scenario occurs for arbitrarily small noise when the dynamics in one basin approaches
   the stable manifold of the saddle point, which guides the dynamics to the
   saddle. Once in the neighborhood of the saddle point, noise may cause
   the switch from one basin to another along the direction of the unstable
   manifold associated with the unstable eigenvalue.


When noise is added to the system, we wish to compute the probability of
escaping from the basin of attraction of attractor $(\vx_a,\vy_a)$. The
asymmetry of the noise between the virtual and real agents is controlled by
$\epsilon$, which scales the noise intensity of ${\vx}$. Computing the
probability of escape  in
the small noise limit implies that we compute  the most likely paths which
cross the  basin boundary of the attractor at the saddle point $(\vx_s,\vy_s)$.
In describing the effect of how noise bleeds into the virtual world from the real
world, we want to measure  noise-induced changes that are large in the dynamics of the state of
${\vx}$ while ${\vy}$ remains approximately stationary; i.e., ${\vy}$
does not change as much as ${\vx}$. Thus, in the presence of noise, we are  
interested in describing how the most likely path develops when ${\vx}_a(K)$
changes its position much more than  ${\vy}_a(K)$. In order to focus on this case, we assume that, when noise and coupling are both 0, ${\vx}_a(0)$ lies in a different part of phase space than ${\vy}_a(0)$. Correspondingly, we also assume that  $||{\vx}_a(K)- {\vx}_s(K)|| \gg ||{\vy}_a(K)- {\vy}_s(K)||$, while $||{\vy}_a(K)- {\vy}_s(K)|| \ll 1$ given that the equilibria depend smoothly on the coupling strength $K$. We note, however, that these assumptions do not affect the general theory, and our methods could be equally well applied if these assumptions were dropped.

\subsection{The Variational Formulation of Noise Induced Escape}

For a given coupling $K$, we wish to determine the path with the maximum probability of noise induced
 switching from the initial attracting state $({\vx}_a(K),{\vy}_a(K))$ to another $({\vx}_b(K),{\vy}_b(K))$, where the
 initial and final states are equilibria of the noise-free versions of Eqs.~\ref{eq:SDE1and2}. Each attractor possesses its own
basin of attraction, and therefore on average, small noise is expected to
induce small fluctuations about the stable equilibria. However, sometimes the
noise will organize itself in such a way  that a large fluctuation occurs, allowing 
escape over the effective energy barrier away from the stable equilibrium. 
If the fluctuation is sufficiently large to bring the system state close to the saddle point,
there is a possibility of switching. Near the saddle point,  depending on the sign of the projection of the local trajectory onto
the unstable manifold of the positive eigenvalue, the
system will approach one or the other attractor. Switching occurs once the
trajectory enters a different basin of attraction from the one where it started.

   We assume the noise intensity $D$ is  much
   smaller than the effective barrier height, and that the scaling on the noise input $\epsilon$ satisfies $0 \le \epsilon < 1$. 
Note that the noise terms $(\xib_\bx,\xib_\by)$ are formally the time derivative
of a Brownian motion, sometimes referred to as white noise \cite{fleming1975deterministic}.

For $D$ sufficiently small, we make the ansatz that the probability distribution of observing  such a large fluctuation  scales exponentially as the inverse of $D$ \cite{FW84,Dykman1990}, 
\begin{equation}\label{rare:event}
\mathcal{P}_{x} = e^{-R/D},
\end{equation} 
where
\begin{equation}\label{rare:event2}
 R(K) =\min_{(\vx,\vy,\boldsymbol{\xi}_x,\boldsymbol{\xi}_y,\boldsymbol{\lambda_1},\boldsymbol{\lambda_2})} \mathcal{R}(\vx,\vy,\boldsymbol{\xi}_x,\boldsymbol{\xi}_y,\boldsymbol{\lambda_1},\boldsymbol{\lambda_2};K),
\end{equation}
and
\begin{widetext}
\begin{multline}
\mathcal{R}(\vx,\vy,\xib_x,\xib_y,\boldsymbol{\lambda_1},\boldsymbol{\lambda_2};K)
= R_{\xib_x} [\xib_x(t)] + R_{\xib_y} [\xib_y(t)]\\ 
+ \int_{-\infty}^{\infty} dt \boldsymbol{\lambda_1}(t)\cdot [
\dot{\vx}(t)-\boldsymbol{f}(\vx(t)) - {\bf h}_1(\vx(t),\vy(t),K) -
\epsilon \boldsymbol{G}_1({\vx}(t)) \xib_x(t)] \\
+ \int_{-\infty}^{\infty} dt \boldsymbol{\lambda_2}(t)\cdot [\dot{\vy}(t) - \boldsymbol{f}(\vy(t)) -
{\bf h}_2(\vx(t),\vy(t),K) - {\boldsymbol G}_2(\vy(t))\xib_y(t)]. 
\label{exponent}
\end{multline}
\end{widetext}
We will see later that the Lagrange multipliers, $\boldsymbol{\lambda}_1,\boldsymbol{\lambda}_2$, also correspond to the conjugate momenta of the equivalent Hamilton-Jacobi formulation of this
problem \footnote{The vector multiplication here is assumed to be an inner
  product.}. Similar to classical mechanics, the exponent
$R$ of Eq. \ref{rare:event} is called the action, and corresponds to the minimizer of the action in the Hamilton-Jacobi formulation which occurs along the optimal path~\cite{feyhib65}. This path
will minimize the integral of Eq.~\ref{exponent}, and is found by setting
the variations along the path $\delta\mathcal{R}$ to zero. The transition rate exponent is proportional to the action, $R$.

When computing the action, the boundary conditions are important, especially
since in general they depend on the parameters of the problem. Therefore, we
suppose that dynamics starts near the attractor $(\vx_a,\vy_a)$. Small
fluctuations will on average remain in the basin of the attractor until at some
point in time, the dynamics hits the saddle point, $(\vx_s,\vy_s)$. Thus, we have
the boundary conditions given by:
\begin{align}
\lim_{t \rightarrow -\infty} ({\vx}(t),{\vy}(t)) &= ({\vx}_a(K),{\vy}_a(K))\\
\lim_{t \rightarrow \infty} ({\vx}(t),{\vy}(t)) &= ({\vx}_s(K),{\vy}_s(K)).
\label{eq:BCS}
\end{align}

To examine the structure of the Hamiltonian governing the large fluctuations,
we take the variational derivative of
$\mathcal{R}(\vx,\vy,\xib_x,\xib_y,\boldsymbol{\lambda_1},\boldsymbol{\lambda_2};K)$
with respect to the noise sources, $\boldsymbol{\xi}_i$ (where $i=\vx,\vy$). Setting the derivative equal to 0 gives
\begin{align}
\boldsymbol{\xi_x} &= 2 \epsilon \boldsymbol{G}_1({\vx}) \boldsymbol{\lambda}_1
\label{eq:lambda1}\\
\boldsymbol{\xi_y} &= 2 \boldsymbol{G}_2({\vy)} \boldsymbol{\lambda}_2.
\label{eq:lambda2}
\end{align}
The full set of  equations of motion is then derived by taking the variational derivatives with respect to the state variables and their corresponding momenta:
\begin{equation} \label{eq:EOM}
\begin{aligned}
\dot{\vx} &= \boldsymbol{f}(\vx) + {\bf h}_1(\vx,\vy,K) + 2 \epsilon^2\boldsymbol{G_1}^2(\bm{x})\bm{\lambda}_1 &&\\
\dot{\bm{\lambda}_1} &= -\epsilon^2 \bm{G_1(x)}{\frac{\partial \bm{G_1}{(\bm{x})}}{\partial \bm{x}}}\bm{\lambda}_1\bm{\lambda}_1
 - \frac{\partial(\bm{f}(\vx)+{\bf h}_1(\vx,\vy,K))}{\partial \bm{x}}\bm{\lambda}_1 \\
&- \frac{\partial({\bf h}_2(\vx,\vy,K))}{\partial
  \bm{x}}\bm{\lambda}_2&&\\
\dot{\vy} &= \boldsymbol{f}(\vy) + {\bf h}_2(\vx,\vy,K) + 2 \boldsymbol{G_2}^2({\bm{y}})\bm{\lambda}_2 &&\\
\dot{\bm{\lambda}_2} &= -\bm{G_2(y)}{\frac{\partial \bm{G_2}{(\bm{y})}}{\partial \bm{y}}}\bm{\lambda}_2\bm{\lambda}_2
-\frac{\partial(\bm{f(y)}+{\bf h}_2(\vx,\vy,K))}{\partial
  \bm{y}}\bm{\lambda}_2\\
 &- \frac{\partial({\bf h}_1(\vx,\vy,K))}{\partial
  \bm{y}}\bm{\lambda}_1. &&
\end{aligned}
\end{equation}

The full Hamiltonian is derived by  substituting  the ansatz in
Eq.~\ref{rare:event} into the appropriate  Fokker-Planck equation and dropping terms of order higher than $1/D$,  which
results in a Hamilton-Jacobi equation with Hamiltonian: 
\begin{align}
\label{hamiltonian} 
H &=  [\epsilon^2
  \boldsymbol{G_1}^2(\boldsymbol{x})\boldsymbol{\lambda}_1]\cdot\boldsymbol{\lambda}_1+
 [\boldsymbol{G_2}^2(\boldsymbol{y})\boldsymbol{\lambda}_2]\cdot\boldsymbol{\lambda}_2\\ \nonumber
&+\boldsymbol{\lambda}_1\cdot[\boldsymbol{f}(\vx) + {\bf h}_1(\vx,\vy,K)]\\ \nonumber
&+\boldsymbol{\lambda}_2\cdot[\boldsymbol{f}(\vy) + {\bf h}_2(\vx,\vy,K)].
\end{align}

One immediate observation from Eq.~\ref{eq:EOM} is that from the conjugate
variables, $({\bm \lambda}_1,{\bm \lambda}_2) \equiv  ({\bm 0},{\bm 0})$ is an
invariant manifold. Moreover, for the system to remain at the equilibria, in
Eq.~\ref{eq:EOM}, the conjugate variables must vanish. (Here, we assume that
multiplicative noise functions do not vanish at the equilibria.) Although the action is
in the exponent of the distribution, the conjugate momenta act as an effective
control
force that pushes the system along a most likely path from the attractor to
the saddle point.   From Eqs.~\ref{eq:lambda1}~and~\ref{eq:lambda2}, it is
therefore  clear
that the noise must  be related to  a large fluctuation governed by the
conjugate variables in the system. Since
at the equilibrium points of  the attractor or saddle, the noise does not
contribute to the exponent of the distribution,  we assume that the
other boundary conditions at equilibrium points for ${\bm \lambda}_i$ are
\begin{align}
\lim_{t \rightarrow \pm \infty} ({\bm \lambda}_1(t),{\bm \lambda}_2(t)) &= ({\bf 0},{\bf 0}).
\label{eq:lambdaBCS}
\end{align}


Locating and computing the most likely, or optimal, path for basin escape
revolves around computing the solution to the two point boundary value problem
consisting of Eqs.~\ref{eq:EOM} and boundary condition Eqs.~\ref{eq:lambdaBCS}~and~\ref{eq:BCS}. However, one must check the local stability of the equilibria
at the boundaries. It can be shown that if the attractor and saddle points in
the deterministic system are hyperbolic,
then the full set of conservative equations of motion will have saddle points
at the boundaries. That is, both the deterministic  attractors and saddles will appear as 
saddles in the Hamiltonian formulation. A fairly
general proof in finite dimensions as well as a useful general method of computing the
solutions for the optimal path  can be found in~\cite{Lindley2013}.

Finally, we note that once we have the optimal path satisfying the variational
problem above, the switching rate from one attractor to the other is given to
logarithmic accuracy by 
\begin{equation}
W = C {\exp}(-\frac{R}{D}),
\end{equation}
where $C$ is a constant and $R$ is given by Eq.~\ref{rare:event2}.

\subsection{Perturbation of Variation}

Because the optimal-path equations are in general nonlinear,
solving them analytically is unrealistic. However, in the case where the
coupling constant $K$ is small, we can use  perturbation theory,
assuming that the variational trajectories remain close to the corresponding trajectories
for $K=0$. Even though the measured perturbation terms will be small, they affect the exponent of the distribution, and since the action is divided by a small intensity,
$D$,   even a small change in the action could have a large effect on the density and mean switching times.

Assuming the terms in the vector field of Eq.~\ref{eq:EOM} are sufficiently
smooth, we suppose the coupling terms $({\bf h}_1(\vx,\vy,K),{\bf h}_2(\vx,\vy,K))$ may be expanded in terms of $K$ as:
\begin{align}
{\bf h}_1(\vx,\vy,K) &= K \hat{{\bf h}}_1(\vx,\vy) + O(K^2)\\
{\bf h}_2(\vx,\vy,K) &= K \hat{{\bf h}}_2(\vx,\vy) + O(K^2).
\label{eq:CK}
\end{align}
Using Eq.~\ref{eq:CK}, we can write, to first order in $K$:
\begin{equation}
\begin{aligned}
\mathcal{R}(\vx,\boldsymbol{\xi},\vy,\boldsymbol{\xi},\boldsymbol{\lambda_1},\boldsymbol{\lambda_2};K)&=
\mathcal{R}_0(\vx,\boldsymbol{\xi},\vy,\boldsymbol{\xi},\boldsymbol{\lambda_1},\boldsymbol{\lambda_2})\\
&+K\mathcal{R}_1(\vx,\boldsymbol{\xi},\vy,\boldsymbol{\xi},\boldsymbol{\lambda_1},\boldsymbol{\lambda_2})
\end{aligned}
\end{equation}
where
\begin{widetext}
\begin{align}
\mathcal{R}_0(\vx,\boldsymbol{\xi_x},\vy,\boldsymbol{\xi_y},\boldsymbol{\lambda_1},\boldsymbol{\lambda_2})
&= R_{\xi_x} [\xib_x(t)] + R_{\xi_y} [\xib_y(t)]\\ 
&+ \int_{-\infty}^{\infty} dt \boldsymbol{\lambda_1}(t)\cdot [
\dot{\vx}(t)-\boldsymbol{f}(\vx(t)) -  
\epsilon \boldsymbol{G}_1({\vx}(t)) {\bf \xi_x}(t)] \\
&+ \int_{-\infty}^{\infty} dt \boldsymbol{\lambda_2}(t)\cdot [\dot{\vy}(t) - \boldsymbol{f}(\vy(t))  - {\boldsymbol G}_2(\vy(t))\boldsymbol{\xi_y}(t)] 
\label{R0}
\end{align}
\end{widetext}
and 
\begin{equation}
\begin{aligned}
\mathcal{R}_1(\vx,\boldsymbol{\xi_x},\vy,\boldsymbol{\xi_x},\boldsymbol{\lambda_1},\boldsymbol{\lambda_2})
&= \\
&- \int_{-\infty}^{\infty} dt [\boldsymbol{\lambda_1}(t)\cdot \hat{{\bf h}}_1(\vx(t),\vy(t))\\
&+ \boldsymbol{\lambda_2}(t)\cdot \hat{{\bf h}}_2(\vx(t),\vy(t))].
\label{R1}
\end{aligned}
\end{equation}

The first order correction to the action can be found by first finding the
solution $(\vx^0,\boldsymbol{\xi_x}^0,\vy^0,\boldsymbol{\xi_y}^0,\boldsymbol{\lambda_1}^0,\boldsymbol{\lambda_2}^0)$
that  minimizes
$\mathcal{R}_0(\vx,\boldsymbol{\xi_x},\vy,\boldsymbol{\xi_y},\boldsymbol{\lambda_1},\boldsymbol{\lambda_2})$.
We then explicitly evaluate the integral in Eq.~\ref{R1} at the
zeroth-order minimization. We note that higher order terms may be found by
applying standard perturbation theory to the equations of motion and boundary
conditions directly, or we may use the general distribution
theory~\cite{Bouchet2016} to get the next order in K, which we do below.

The Hamiltonian for the variation of the action $\mathcal{R}_0$ of the uncoupled system is given by
\begin{equation}
\begin{aligned}
H^0 &=  \epsilon^2
  [\boldsymbol{G_1}^2(\boldsymbol{x}^0)\boldsymbol{\lambda}^0_1]\cdot\boldsymbol{\lambda}^0_1+
  [\boldsymbol{G_2}^2(\boldsymbol{y}^0)\boldsymbol{\lambda}^0_2]\cdot\boldsymbol{\lambda}^0_2\\ 
&+\boldsymbol{\lambda}^0_1\cdot \boldsymbol{f}(\vx^0 )+\boldsymbol{\lambda}^0_2\cdot\boldsymbol{f}(\vy^0).
\label{H0}
\end{aligned}
\end{equation}
The structure of Eq.~\ref{H0} is such that the total action is just the sum of the
action of the $\vx$ and $\vy$ variables since $K=0$. In addition the initial
and final states for $K=0$ are given for the attractor $(\vx^0_a,\vy^0_a)$ and
saddle $(\vx^0_s,\vy^0_s)$. Since we are interested in moving $\vx$ through a
large fluctuation while holding $\vy$ approximately constant, when uncoupled
the initial states satisfy $\vx^0_a \ne \vy^0_a$, while $\vy^0_s = \vy^0_a$, the
latter assuming no movement in $\vy$.

The effect of the noise reducing parameter on the action is now evident from
the equations of motion derived from Eq.~\ref{H0}. The total action is just the
sum of the $\vx$ and $\vy$ actions, ${\cal R}^0[\vx],{\cal R}^0[\vy]$, respectively. Moreover, 
since there is no movement in
$\vy$,  ${\cal R}^0[\vy] \equiv 0$. Assuming the multiplicative noise term
is non-singular, the resulting uncoupled action is
therefore given by 
\begin{equation}
\label{R0x}
{\cal R}^0[\vx] = -\frac{1}{\epsilon^2} {\int_{\vx_a}^{\vx_s}} [{{\bm
    G}_1^2}]^{-1} {\bm f}(\vx) d \vx.
\end{equation}
The expected effect of the parameter $\epsilon$ is evident, in that the action
scales as $\frac{1}{\epsilon^2}$. Coupled with the fact that the action is in the
exponent of the distribution means that the exponent should scale as
$\frac{1}{\epsilon^2 D}$, which will make the probability of $\vx$ transitioning
through a large fluctuation conditioned on $\vy$ staying approximately constant
a very rare event. 

Notice that we can also consider how $\vy$ switches while keeping $\vx$ approximately constant, by changing the boundary conditions. In this case the switching rate is much higher
since the exponent of the switching rate scales as $\frac{1}{D}$.

To see how such a rare event explicitly comes about, we consider the following
generic bi-stable situation.

\section{A model example of mixed reality noise induced perturbations}
\label{sec:AModelExampleOfMixedRealityNoiseInducedPerturbations}

{For clarity, we now give an example of noise-induced switching in a generic coupled
system where the individual components are affected by different scales of noise.
Consider} two coupled
particles interacting in a double-well potential $U(x)$. One particle represents
a simulated robotic agent, while the other represents the real-world robot
that interacts with the simulation. The two-particle system is used because it
is sufficiently complex to illustrate our argument, while remaining simple
enough to be understood analytically. Our approach follows the general theory of
switching in the previous section, but for purposes of
analysis we consider the following symmetric double-well potential: 
\begin{equation}
 U(x) = \frac{x^4}{4} - \frac{x^2}{2}.
\end{equation}
In the absence of coupling, the resulting motion of a single particle is described by 
 $\frac{\rd x}{\rd t} = f(x)=-\frac{dU(x)}{dx}$. Now
suppose that the $(x,y)$ particles are coupled with a spring potential~\cite{szwaykowska2016collective}, and a white
Gaussian noise $\xi_x,\xi_y$ is assumed to act on each particle independently. Let $x$ and $y$ denote the positions of particles 1 and 2, respectively. Their equations of motion are then:
\begin{subequations} \label{eq:fullsys}
\begin{align}
 \dot{x} &= f(x) - K\cdot(x-y) +  \epsilon \xi_x\\
 \dot{y} &= f(y) + K\cdot(x-y) + \xi_y.
\end{align}
\end{subequations}
We assume that $E[\xi_x(s)\xi_y(t)] = 2D\delta(t-s)\delta(x-y)$, where $D \ll 1$ is the noise
intensity, and $\epsilon$ the satisfies the hypotheses in the previous section.

\subsection{The deterministic picture}

\begin{figure}[h]
 \centering
 \includegraphics[width=0.5\textwidth]{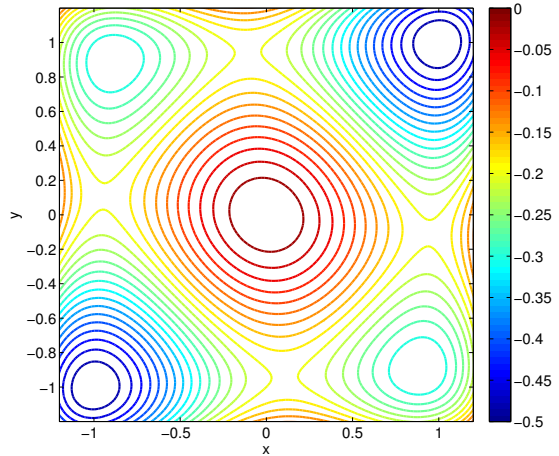}
 \caption{(Color) A contour plot of the potential function $V(x,y;0.1)$.}
 \label{fig:2dpotential}
\end{figure}

Consider the noise-free system obtained by setting $\xi_x \equiv 0, \xi_y\equiv 0$ in (\ref{eq:fullsys}).
The system has an effective potential $V(x,y;K)$ given by:
\begin{equation}
\begin{split}
V(x,y;K) = &-\frac{1}{2}\,{x}^{2}+\frac{1}{4}\,{x}^{4}-\frac{1}{2}\,{y}^{2}+\frac{1}{4}\,{y}^{4}\\
& +\frac{1}{2}\,K \left( x-y
 \right) ^{2}
\end{split}
\label{eq:2dpotential}.
\end{equation}
The topology of the equilibria for $K=0.1$ is pictured in
Fig.~\ref{fig:2dpotential}. 

The system has stable
equilibria at $(x,y) = (-1,-1)$ and $(1,1)$. The equilibrium solution $(x,y) = (0,0)$ is unstable for $K<1/2$ and a saddle
point for $K>1/2$. For $K \leq 1/2$, the symmetric configuration about $0$, with $(x,y) = (\pm\sqrt{1-2K},\mp\sqrt{1-2K})$, is stable for
$K \in [0,1/3)$ and a saddle for $K \in (1/3,1/2]$. As $K \rightarrow 1/2^-$, this solution approaches the unstable equilibrium at $(0,0)$. The solutions collide at $K=1/2$, resulting in a saddle point at $(0,0)$.

\begin{figure}[h]
 \centering
 \includegraphics[width=0.5\textwidth]{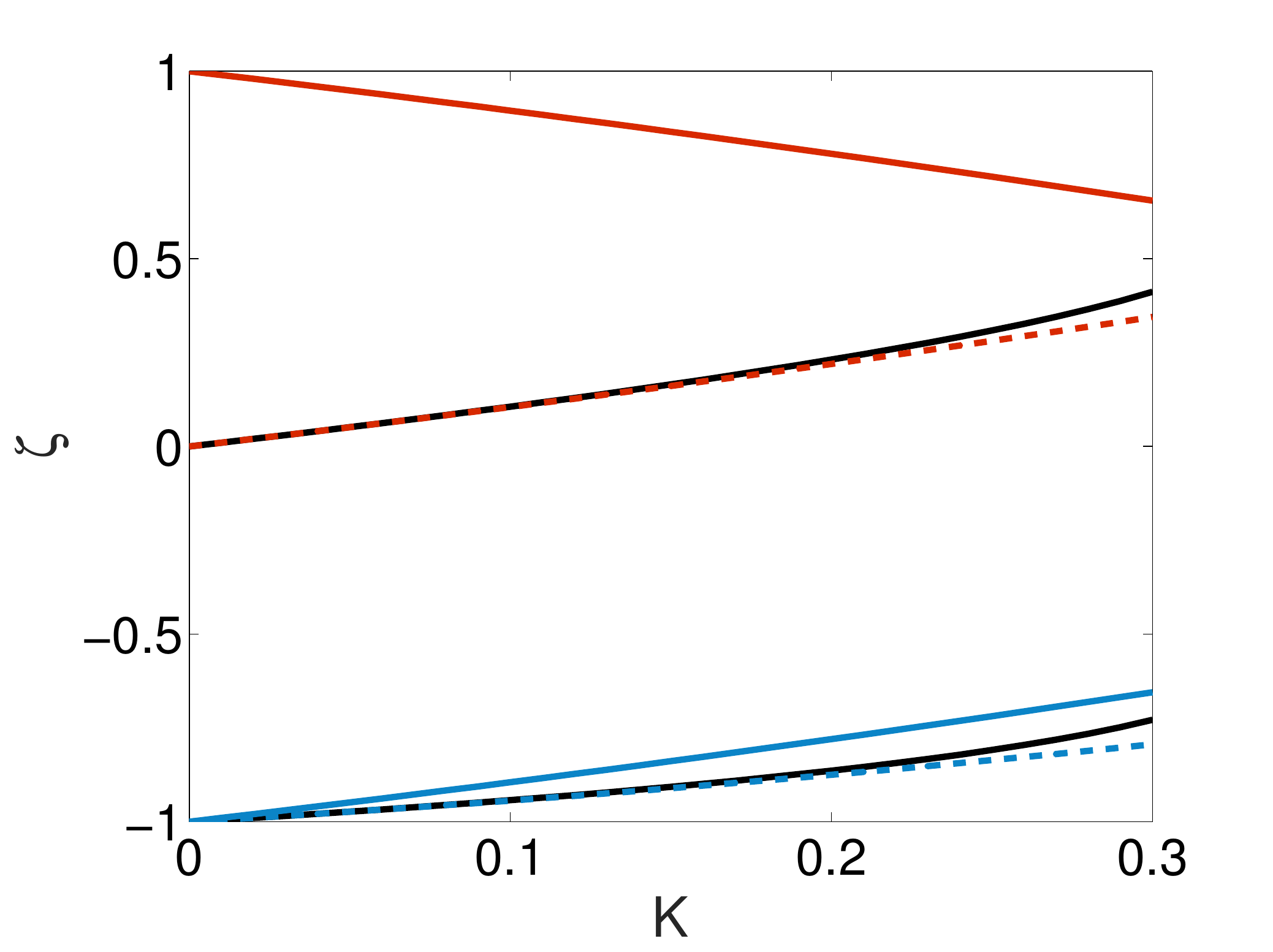}
 \caption{
   (Color) Values of zeroes of the deterministic vector field as a function of the coupling strength $K$. The solid black lines denotes a solutions to the exact expression in Eq.~\ref{eq:zeta}, showing the location of a saddle point $(x_s, y_s)$ of the system. Dashed lines denote the asymptotic approximate of the saddle point location for $K \ll 1$. Solid colored lines denote the location of the attractor $(x_a, y_a)$. Blue and red are used to denote the positions of $\bar{x}$ and $\bar{y}$, respectively, where $\bar{x} = x_a$ or $x_s$ and $\bar{y} = y_a$ or $y_s$.}
 \label{fig:zetaVk}
\end{figure}

The system has 4 additional equilibria, defined by
\begin{align}
\label{eq:zeta}
 (x,y) = (\zeta, \frac{\zeta}{K}(\zeta^2 + K - 1))
\end{align}
where $\zeta$ is a root of 
\begin{equation}
\zeta^4 + (K-1)\zeta^2 + K^2 = 0.
\label{zetaroot}
\end{equation}
 Solutions exist for $K \in [0,1/3]$; the corresponding
equilibria are saddle points. 
A plot of the equilibria for this system for different $K$ is shown in Fig. \ref{fig:zetaVk}.

\subsection{Switching}


When adding noise into the system, it is possible to observe noise-induced switching between stable equilibria of the
noise-free system. Here we will derive most-likely noise-induced switching
paths starting from the stable symmetric configuration where the particles are
located in separate basins; $x$ then experiences a large
fluctuation and transitions to the basin occupied by $y$. In
the small-noise limit, the most likely path passes through a saddle point of
the noise-free system. In the following analysis, we therefore compute the
optimal switching path from the initial system configuration to the saddle
point; the remaining transition from the saddle to the final stable
configuration occurs much more rapidly, since it is dominated by the
deterministic dynamics.

\begin{figure}[h]
  \centering
  \includegraphics[width=0.4\textwidth]{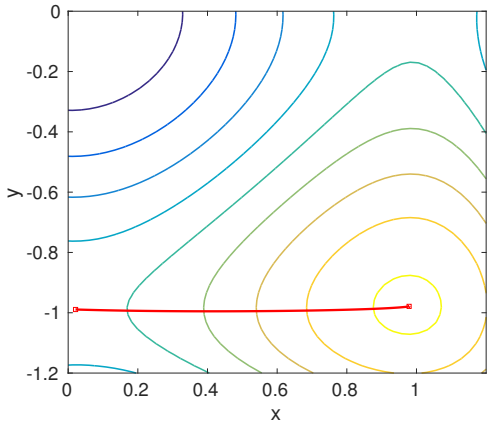}
  \caption{(Color) The potential function Eq.~\ref{eq:2dpotential} for $K=0.0209$ for the
    zero noise case.  Overlaid is the predicted
    optimal path (red line) computed when $\epsilon=\frac{1}{2}$. For the
    boundary conditions chosen, $x$ starts near 1 and goes through a large fluctuation, while $y$
    remains approximately stationary near -1.}
  \label{fig:2dcontourwithpath}
\end{figure}


For sufficiently small noise intensity $D$, the switching dynamics can be described using the
Hamiltonian formulation of Eq.~\ref{eq:EOM}, where we extend the system to 4 dimensions by adding in conjugate momenta ($\lambda_1$ and
$\lambda_2$), and set the multiplicative noise terms to the identity:
\begin{subequations}
\label{eq:exeom}
 \begin{align}
 \dot{x} &= f(x) - K\cdot(x-y) + 2 \epsilon^2 \lambda_1 \\
 \dot{y} &= f(y) + K\cdot(x-y) + 2 \lambda_2 \\
 \dot{\lambda}_1 &= -(f'(x) - K)\lambda_1 - K\lambda_2 \\
 \dot{\lambda}_2 &= -(f'(y) - K)\lambda_2 - K\lambda_1,
 \end{align}
\end{subequations}
with corresponding Hamiltonian
\begin{align} \label{eq:exham}
 \H (x,y,\lambda_1,\lambda_2) &= [f(x) - K\cdot(x-y)]\lambda_1\\ \nonumber
 &+[f(y) + K\cdot(x-y)]\lambda_2 + \lambda_2^2 +  \epsilon^2 \lambda_1^2. 
\end{align}
Note that $\H(x^*,y^*,0,0) = 0$ for all $(x^*,y^*)$ in the set of equilibria of (\ref{eq:fullsys}). Since $\H$ is
time-invariant, optimal switching paths between equilibria are required to satisfy a two-point boundary problem  on the
zero level sets of $\H = 0$ in order to compute the action.

We use the numerical approach described in
\cite{Lindley2013} to compute the optimal path starting at $(x_a, y_a) = (\sqrt{1-2K}, -\sqrt{1-2K})$ for $t \rightarrow -\infty$ and passing through the saddle point given by
$(x_s, y_s) = (\zeta, \frac{\zeta}{K}(\zeta^2 + K - 1)) \approx (K+K^2/2,-1 + 2K + 5 K^2/8)$ as $t \rightarrow \infty$, where $1/\sqrt{3} < \zeta < 1$, and $K \ll 1$. An example of such a path is shown in Fig.~\ref{fig:2dcontourwithpath}.

In Eqs.~\ref{eq:exeom}, consider the limit $K \rightarrow 0$. The particle motions are uncoupled, and the situation is equivalent to a
single-particle switching problem. In this case, it is possible to find an
analytic solution in time explicitly, and make use of the general perturbation
theory. From Eq.~\ref{R0x}, we know that for non-zero $\epsilon$, the zeroth
order term of the action  scales inversely with $\epsilon^2$, and in fact is
given by 
\begin{equation}
R^0 =  
\frac{1}{4 \epsilon^2},
\end{equation}
where we have used the fact that from the Hamiltonian, the optimal path when
$K = 0$ is
given explicitly by $\lambda_1^0 = -\frac{1}{\epsilon^2} f(x^0)$.

To get the first order corrections, we need the solution to the two point value
problem along the zeroth order optimal path as a function of time:

\begin{subequations}
 \begin{align}
  x^0(t) &= \frac{1}{\sqrt{1 + e^{2t}}} \\
  y^0(t) &= -1 \\
  \lambda_1^0(t) &= -{\frac {{{\rm e}^{2\,t}}}{ \left( 1+{{\rm e}^{2\,t}} \right) ^{3/2}{\epsilon}^{2}}}\\
  \lambda_2^0(t) &= 0
 \end{align}
\end{subequations}

Notice that as $t \rightarrow \pm \infty$, we have the following boundary
conditions satisfied for $(x^0(t),\lambda_1^0(t))$ while holding $(y^0 \equiv
-1,\lambda_2^0 = 0)$ constant:
\begin{subequations}
\begin{align}
  \lim_ {t \rightarrow -\infty} &      & \lim_ {t \rightarrow \infty}  \nonumber  \\
  x^0(t) &\rightarrow  x_a^0=1 & x^0(t) &\rightarrow  x_s^0=0  \\
\lambda_1^0(t) &\rightarrow 0   &\lambda_1^0(t) &\rightarrow 0.
\end{align}
\end{subequations}
Using the zeroth order time series in the first order expression of the action gives, to linear order in $K$:
\begin{equation}
{\cal R} = \frac{1}{4 \epsilon^2} - K \frac{3}{2\epsilon^2}.
\end{equation} 


\subsection{Second order effects}
\begin{figure}[thb]
\centering
\includegraphics[width=0.5\textwidth]{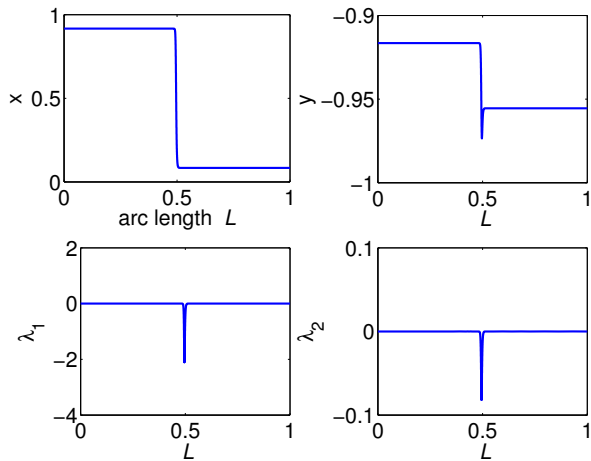}
\caption{(Color) Trajectory projections of the optimal path where time is rescaled to
  be  arc length, {\it L}, along the trajectory. Shown are trajectories for
  $x,y$ and their conjugate variables $\lambda_1,\lambda_2$. The parameters
  used are $\epsilon=0.5, K=0.08$.}
\label{fig:b0p5}
\end{figure}

We can get the second order effects of the coupling strength $K$ on the action by considering the potential function of
Eq.~\ref{eq:2dpotential}, and using the general results of computing the
probability of escape for Gaussian noise in \cite{Bouchet2016}. However, the
approach here is one that will be problem-specific.  We choose to formally 
examine the Hamiltonian in Eq.~\ref{eq:exham}, and notice that  $y$ 
and its conjugate momenta remain approximately near the attractor. Therefore,
we use the asymptotic expression of the attractor and saddle, in the limits of
Eq.~\ref{R1}.
\begin{equation}
\begin{aligned}
\mathcal{R}_1 &= \int_{-\infty}^{\infty} dt [\boldsymbol{\lambda_1}(t)\cdot \hat{{\bf h}}_1(\vx(t),\vy(t))\\
&\quad\quad\quad + \boldsymbol{\lambda_2}(t)\cdot \hat{{\bf h}}_2(\vx(t),\vy(t))]\\
&= -\int_{-\infty}^{\infty} dt
\frac{f(x_{0}(t)}{\epsilon^2}(x_{0}(t)-y_{0}(t))\\
&= \frac{1}{\epsilon^2} \int_{x_{a}(K)}^{x_{s}(K)} dx_{0}(x_{0}+1)
\end{aligned}
\label{Eq:secondorder}
\end{equation}
Using the asymptotic expressions for the attractor and saddle for $x_{0}$, 
expanding for small $K$, and collecting terms, we have
\begin{equation}
\label{eq:action_asymp}
{\cal R} \approx \frac{1}{4 \epsilon^{2}}-\frac{3K}{2\epsilon^{2}}+\frac{2{K}^{2}}{\epsilon^{2}}.
\end{equation} 
An example of the optimal path projections is given in Fig.~\ref{fig:b0p5}
for moderate noise reduction ($\epsilon=0.5$) and small coupling $K$. Notice that
in the figure, $(x(t),y(t))$ spend most of their time near the equilibria
specified at the boundaries. In addition, $x(t)$ traverses a distance of order unity
when it switches from the attractor
to the saddle point, while deviations of $y(t)$ from the equilibrium position are only of order $K$. Therefore, even though the 
scaled reduction of the noise parameter is small, the noise transmitted to $x$
has a very strong effect through the coupling.

\begin{figure}[thb]
\centering
\includegraphics[width=0.5\textwidth]{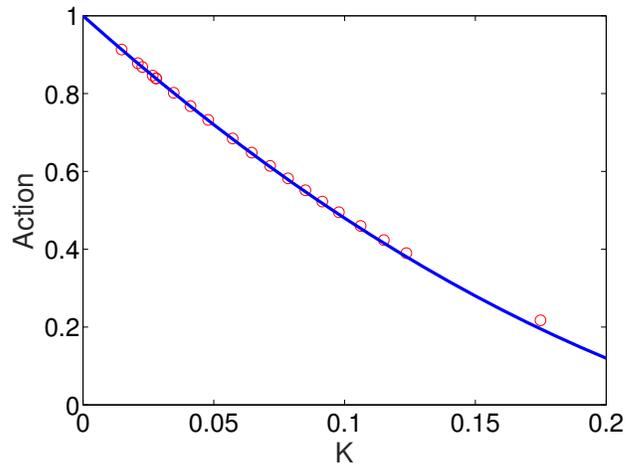}
\caption{(Color) The numerically computed action (red circles)  plotted as
  a function of coupling parameter $K$, for $\epsilon = 0.5$. The asymptotic result for small $K$ obtained using
Eq.~\ref{eq:action_asymp} is shown by the blue curve.}
\label{fig:ActionAsymp}
\end{figure}

Using the theory for the action, Fig.~\ref{fig:ActionAsymp} shows how it
scales as a function of $K$ when $\epsilon = 0.5$. Along with the
numerically computed action are the results from the asymptotic analysis for small
coupling using Eq.~\ref{eq:action_asymp}. Notice that for $K < 0.2$, 
 the agreement is good, and improves as $K$ gets smaller.

\begin{figure}
\centering
\includegraphics[width=0.5\textwidth]{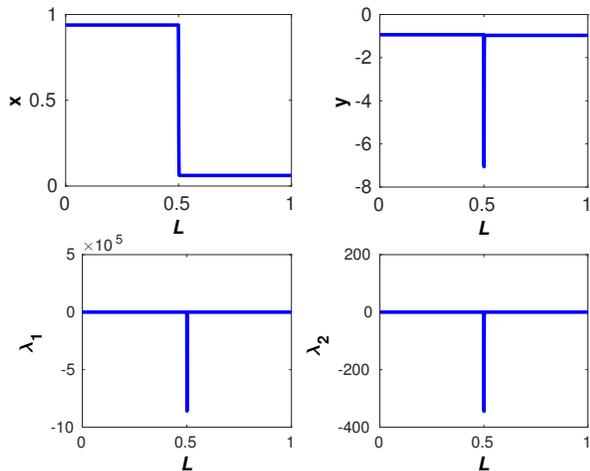}
\caption{(Color) Optimal switching path for the system in (\ref{eq:exeom}),
  with $K=0.0595$ and $\epsilon=0$ to machine precision. }
\label{fig:K0p1b0}
\end{figure}

One of the interesting facets of the problem  occurs when there is noise only on the $y$ component. This situation occurs when  $\epsilon$ approaches zero. Although asymptotically the action is seen to approach
$\infty$ as $\epsilon$ approaches zero, since the system is coupled it is possible to compute an optimal path  conditioned on  that large
fluctuations occur only in $x$.
Using the results for finite $\epsilon$ as an initial guess, we  use continuation to decrease $\epsilon$ to $0$, and obtain the
optimal path for switching in the coupled system with noise acting only on
particle $y$ (see Fig. \ref{fig:K0p1b0}), where the coupling constant is
relatively small; i.e. $K \approx 0.06$. 
The action along the optimal path is on the order of $10^5$, which indicates
that switching would be an extremely rare event. In this case we do observe a relatively large change in $y$ which is on the order of unity rather than $K$, but $y$ does spend most its time near its equilibrium. 

\begin{figure}[thb]
\centering
\includegraphics[width=0.5\textwidth]{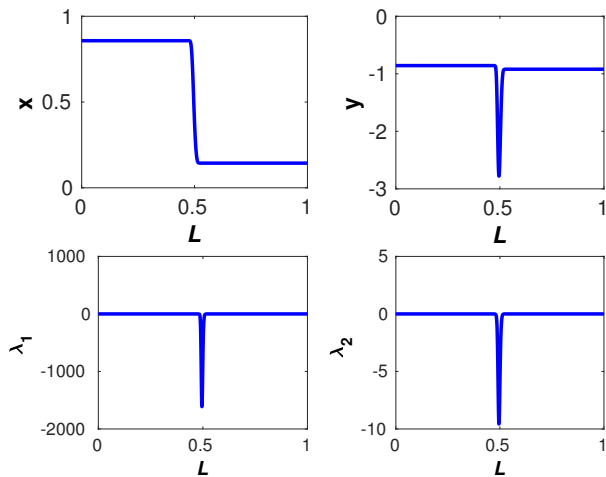}
\caption{\label{fig:K0p27} (Color) Trajectory projections of the optimal path where time is rescaled to
  be  arc length, {\it L}, along the trajectory. Shown are trajectories for
  $x,y$ and their conjugate variables $\lambda_1,\lambda_2$. The parameters
  used are $\epsilon=0.0, K=0.1324$.}

\end{figure}

The interaction of the coupling and noise induced forces is key in determining the switching times for the system. Increasing the coupling $K$ by an order of magnitude results in a drastic change in
the values of the conjugate variables along the optimal switching path, as shown in Fig.~\ref{fig:K0p1b0}-\ref{fig:K0p27}. Here we see that in the system with increased $K$ (Fig.~\ref{fig:K0p27}), both $x$ and $y$ still undergo a change of order unity; however, the values of the  conjugate variables $\lambda_1 , \lambda_2$, have been reduced by several orders of magnitude. The action is therefore much smaller (${\cal R} \approx 500$ compared to ${\cal R} \approx 5.7 \cdot 10^5$ with weak coupling), implying a much shorter switching time. 



\begin{figure}
\centering
\includegraphics[width=0.5\textwidth]{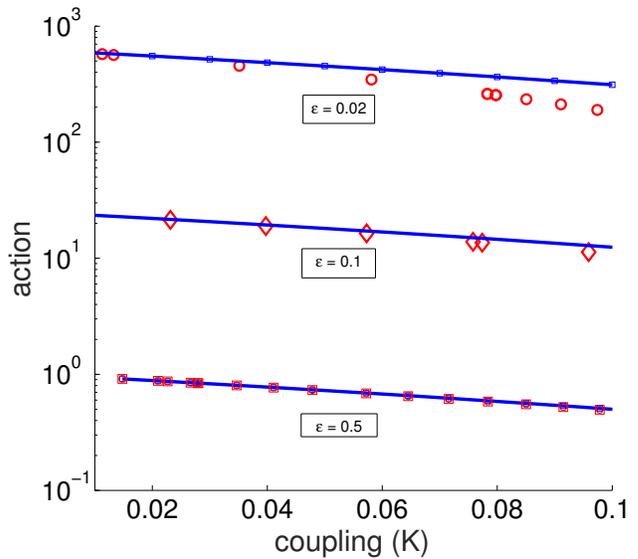}
\caption{(Color) A plot showing the numerically computed log of the action and the asymptotic approximation (Eq.~\ref{eq:action_asymp})
  as a function of coupling strength, $K$. Shown are results for $\epsilon =
  0.5,0.1,0.02$. The computed action is shown in red, while the asymptotic expression is depicted by the blue lines.}
\label{fig:composite}
\end{figure}

The effect of coupling strength on action along the optimal path is shown in Fig.~\ref{fig:composite}, for different values of $\epsilon$. We observe that the range of $K$ for which the asymptotic prediction ($K \ll 1$) of the action holds decreases as $\epsilon$ decreases.

\subsection{Monte Carlo Simulation}

We consider the problem of switching in Eqs.~\ref{eq:fullsys} where the
asymmetry in noise intensity between two coupled systems is governed by the parameter 
 $\epsilon$. Using the Milstein method for numerical solution of stochastic differential equations (SDEs),
we implement a Monte Carlo scheme  to
compute the mean time for the $x$ variable to switch while the $y$ variable
remains in its basin, given that the particles start in different basins of attraction. 
That is, we compute the mean time it takes for  $x$ to transition from
$x(0)=x_a(K)$ to  the saddle point $x(T)=x_s(K)$. 

We first check the existence of an exponential distribution of times by
computing the switching time as a function of the inverse noise intensity for
various values of $\epsilon$ and $K$.  From the ansatz that the mean switching
time exponents are proportional to ${\cal R}/D$, we plot the log of the mean
switching time as a function of $1/D$, where the slope should be the action
evaluated at the parameters of $\epsilon,K$.

\begin{figure}
\centering
\includegraphics[width=0.5\textwidth]{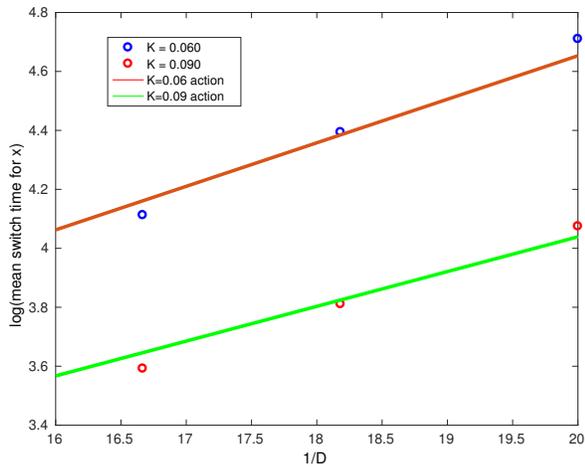}
\caption{(Color) Plotted is the mean switching time for $x$ to go from the attractor
  to the saddle point while $y$ remains in its basin of attraction, as a function of inverse noise intensity ($1/D$). Here
  $\epsilon = 0.75$. Results are shown for $K = 0.060$ and $0.090$. The solid lines show theoretical values computed using Eq.~\ref{eq:exeom}, while values obtained from Monte Carlo simulations are depicted by circles.}
\label{fig:xswitchingvD}
\end{figure}

\begin{figure}
\centering
\includegraphics[width=0.5\textwidth]{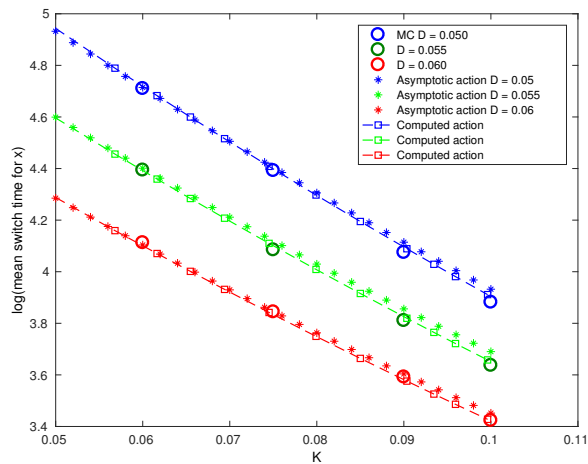}
\caption{(Color) Plotted is the mean switching time for $x$ as a function of coupling strength $K$. Here
  $\epsilon = 0.75$. Results for $D = 0.050$, $0.055$, and $0.060$ are shown in blue, green, and red, respectively. Values obtained from Monte Carlo simulations are plotted as circles; values computed using the asymptotic theory (Eq.~\ref{eq:action_asymp}) are plotted as stars; and true values for the action are plotted as dashed lines. Since the prefactor is not directly computed in the asymptotic calculations of action, the plotted values are shifted to coincide
  with the first data point. Note that the asymptotic values deviate from the true values for large K.}
\label{fig:xswitchingvK}
\end{figure}

We can see how the asymptotic theory holds as a function of $K$ by comparing
it with the mean switching times in Figs.~\ref{fig:xswitchingvD}-\ref{fig:xswitchingvK}. For small
$K$, the theory holds up quite well for a range of noise intensities (where noise intensity is small compared to the barrier height), and over sufficiently large range of $K$. 



\section{Discussion and Conclusion}
\label{sec:DiscussionAndConclusions}


In this paper, we addressed the problem of how coupling can enhance noise-induced switching in systems with highly asymmetric noise characteristics. As a motivating example, we considered a simple mixed-reality experiment in which a virtual system with very low or zero noise is coupled with a noisy real system. We showed that the effect of coupling was sufficient to
cause the virtual dynamics to undergo a large fluctuation while the real
dynamics, which was driven by larger noise intensity, remained 
quiescent. It was very natural to take a variational approach to describing
such a large fluctuation, and although it was applied to Gaussian noise,
the same approach can
be extended to more general noise sources \cite{Billings2010}. 

Using the variational approach, we generated a Hamiltonian two-point boundary value
problem with asymmetric driving representing the effect of the heterogeneous
noise sources. We used scaling parameter $\epsilon \in [0,1]$ to quantify the asymmetry in the noise. The solution to the Hamiltonian equations generates the optimal switching path, which in
turn can be used to predict mean switching rates.

We focused on the case where a large fluctuation occurs in the low noise system, while the system with higher intensity noise remains near its equilibrium point. Note that, because of the asymmetry in noise levels, the probability of a large transition in the high noise system occurring before the fluctuation in the low noise system is very high. Thus having the low noise system transition first is an extremely rare event.


We illustrated the general theory using a general model of a pair of coupled particles in a bi-stable potential. This example was inspired by bistable behaviors predicted for a mixed-reality system of swarming agents \cite{szwaykowska2016collective}. We quantified the action as a
function of coupling strength over a range of scaling values $\epsilon$, revealing an excellent comparison between asymptotic  theory and numerical solutions of the optimal paths. However, we note that, for very small values of $\epsilon$, the asymptotic theory diverges from the true action for even moderate values of $K$. This is a result of two small parameters in the approximation; higher-order corrections may need to be included in Eq.~\ref{eq:action_asymp}. We also quantified the mean escape times  in terms of parameters $\epsilon$ and $D$, again with excellent agreement between simulation and theory for the log of the mean switching time.

We computed the paths as the noise scaling parameter $\epsilon$ approaches zero, so that the probability of extremely rare events is
governed by coupling strength alone. That is, the noise is only transmitted
through the coupling terms. The asymptotic theory predicts an logarithmic
exponent of the probability of virtual switching given that the real dynamics
exhibits only small fluctuations, where the exponent scales as
$1/\epsilon^2$. Although extremely rare, the switching is still observed when
$\epsilon \rightarrow 0$ and
coupling $K$ is sufficiently large~\cite{Note3}, as we have shown in Fig.~\ref{fig:K0p27}.

The physical interpretation of the transmitted noise induced large fluctuation
is that the coupling also acts as an effective force along with the effective
stochastic momenta to enhance the observation of an extremely rare event. The 
coupling used in the generic example is similar to the couplings found in many physical
systems, including the swarm experiment we described.   Since our theory is generic, it predicts that such noise transmitted fluctuations should
appear in many coupled systems, including mixed-reality  situations, where the noise intensities
are highly skewed.


\section{Acknowledgments}
The authors gratefully acknowledge the Office of Naval Research for their
support under  N0001412WX20083, and support of the NRL Base Research Program
N0001412WX30002. KS was a National Research  post doctoral fellow while
performing the research. We acknowledge useful conversations with Ani Hsieh,
Luis Mier, 
and Brandon Lindley about early versions of the research, and Jason Hindes for
an initial reading of the manuscript.

%

\end{document}